\documentclass[12pt,a4paper]{article}%
\usepackage{amsfonts}
\usepackage{amsmath}
\usepackage{amssymb}
\usepackage{graphicx}%
\providecommand{\U}[1]{\protect\rule{.1in}{.1in}}

\begin{document}

\title{Gravity Cannot Cure Quantum Mechanics of its Malady of the Collapse of the Wavefunction}
\author{C. S. Unnikrishnan$^{1}$ and George T. Gillies$^{2}$\\$^{1}$\textit{Tata Institute of Fundamental Research, }\\\textit{\ Homi Bhabha Road, Mumbai - 400 005, India}\\$^{2}$\textit{School of Engineering and Applied Science,}\\\textit{University of Virginia}, \textit{Charlottesville, VA 22904-4746, USA}\\E-mail address: $^{1}$unni@tifr.res.in, $^{2}$gtg@virginia.edu\\}
\date{This essay received an `honorable mention' in the Gravity Research Foundation
2021 Awards for Essays on Gravitation.}
\maketitle

\begin{abstract}
The speculation that gravity is the key to solving the quantum measurement
problem has been alive for decades, without any convincing demonstration of a
solution. One necessary factor in the relevant proposals is that the
gravitational energy of mutual interaction, which scales quadratically with
the mass, facilitates the spontaneous collapse of the wavefunctions in
spatially separated superpositions. Relying on a simple physical input from
electrodynamics, supported by robust first principle calculations, we show
that the speculations connecting gravity and the hypothetical spontaneous
collapse of the wavefunction are inconsistent and not tenable. The result
suggests that the gravitational solution to the problem of the collapse of the
wavefunction be put to rest.

\end{abstract}

Gravity is the most familiar of fundamental interactions, and its theoretical
understanding has evolved and matured over 300 years. With the General Theory
of Relativity, we have an encompassing theory, applicable from the laboratory
to the largest scale imaginable. Yet, we are not sure about its unconditional
applicability in the smallest of physical scales. The reasons are two-fold.
Since gravity is relatively very weak, experimental indications are difficult
come by from the physics of small masses and microscopic distances. More
importantly, the dynamics at the atomic scales are dictated by the theory of
quantum mechanics, which is known to have issues of compatibility with the
theory of gravity.

Apart from the vigorous efforts to find the unified terrain of gravity and
quantum mechanics, in a future theory of quantum gravity, there is another
important research front that hopes to combine aspects from the two theories.
This is to invoke the macroscopic `classical' theory of gravity to solve the
long-standing vexing problems of quantum mechanics. The situation is somewhat
paradoxical; on the one hand it is assumed that the classical theory of
gravity is definitely incompatible with quantum mechanics, demanding
fundamental modifications in the classical theory of General Relativity. But
on the other hand, one is appealing to even the Newtonian features of the
gravitational interaction to cure the age old maladies of quantum mechanics.
\emph{The primary unsolved problem of the quantum theory is the quantum
measurement problem, which can be traced to its `birth defect' of the collapse
of the wavefunction} \cite{Leggett,Weinberg}. The quantum measurement problem
arises in the collapse of the entangled wavefunction of one microscopic and
another macroscopic system. However, we do not even understand how a quantum
state characterized by a simple superposition of two spatially separated
wavefunctions reduces to one of the two, to faithfully represent the factually
observed situation, in each trial of a quantum mechanical experiment. A
related open issue is how is that we are able to prepare such superpositions
for microscopic physical systems whereas it becomes progressively difficult as
the mass and size of matter increase, eventually becoming impossible in the
case of macroscopic objects.

The proposals that invoke the gravitational interaction as a universal `agent'
to facilitate the spontaneous collapse of a spatial superposition of
wavefunctions have been discussed for several decades now
\cite{Diosi1,Karoly,Pen1,Diosi2,Pen2}. The quantitatively detailed schemes are
associated with the independent proposals from L. Di\'{o}si and R. Penrose,
though often they are collectively called the Di\'{o}si-Penrose (D-P) models.
The difference is in the exact theoretical basis for invoking exclusively
gravity to induce the spontaneous collapse. While the Di\'{o}si mechanism is
largely Newtonian, Penrose bases his reasoning on the Einsteinian association
of gravity with the very properties of space and time where the quantum
dynamics happens \cite{Pen2}.%
\begin{figure}
[ptb]
\begin{center}
\includegraphics[
height=1.8348in,
width=3.1075in
]%
{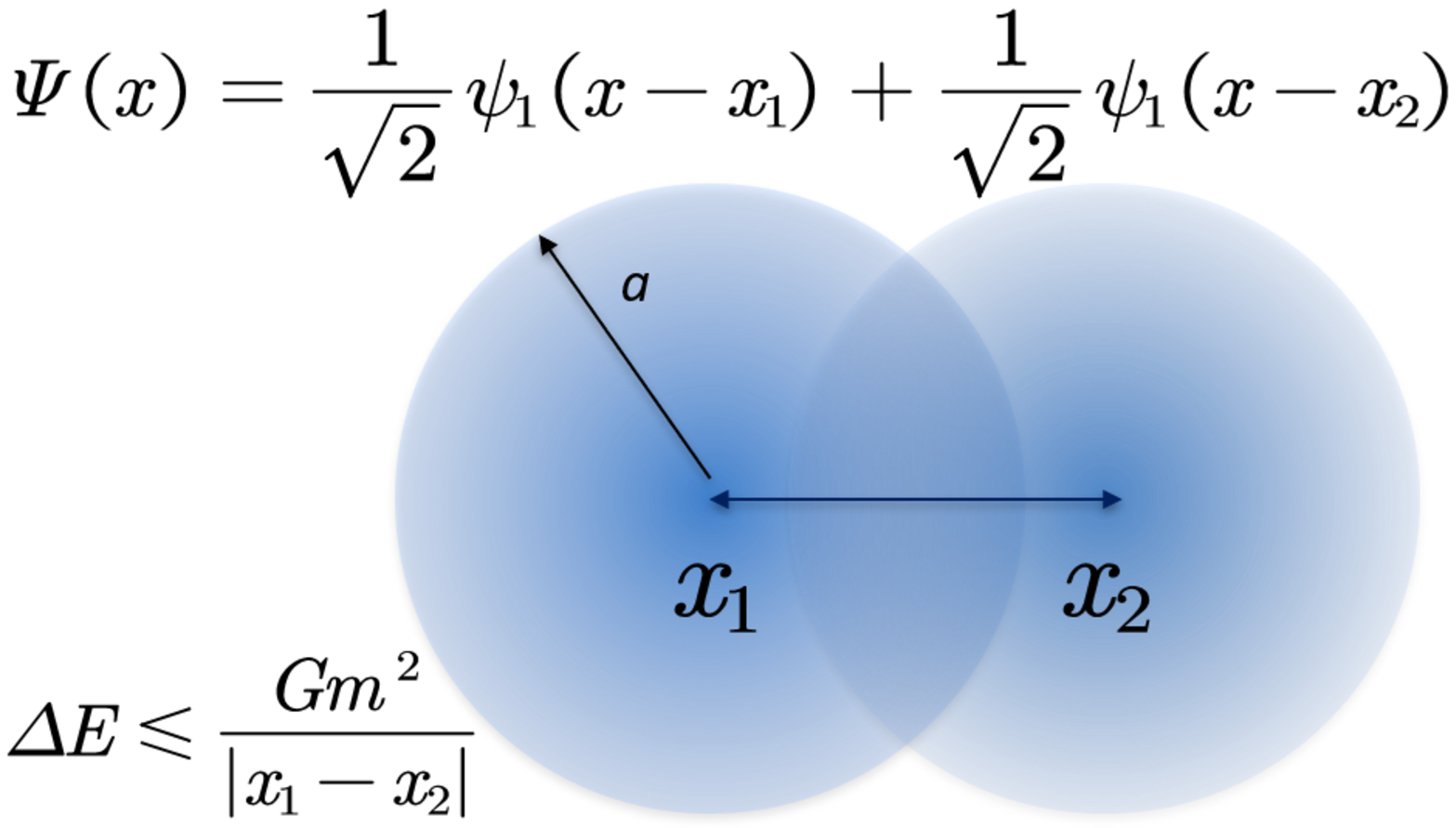}%
\caption{In the models for the gravity-induced spontaneous collapse of the
superpositions of quantum mechanical wavefunctions, as in the proposals by
Di\'{o}si and Penrose, a spatial superposition is assumed to have an
ontological counterpart in real space. Hence, the interaction energy of the
two parts of the wavefunction is well defined. However, the hitherto
overlooked physical aspect of much stronger interactions that directly
influence the dynamics, like the electromagnetic interaction when the particle
is charged, demolishes the entire scheme as a thoroughly nonviable and
inconsistent proposition.}%
\label{super}%
\end{center}
\end{figure}

However, the quantitative details in both mechanisms are very similar. This is
what is important for our analysis here, which eventually reliably rules out
gravity as responsible for the collapse of the quantum superpositions of
wavefunctions. The case under consideration is represented by the simple
superposition like
\begin{equation}
\Psi=a\psi(x-x_{1})+b\psi(x-x_{2})
\end{equation}
that represents the wavefunction of a single particle of mass $m$. The
functions $\psi(x-x_{1})$ and $\psi(x-x_{21})$ could be Gaussian functions
centred on the coordinates $x_{1}$ and $x_{2}$, with their separation
$\left\vert x_{1}-x_{2}\right\vert $ even larger than the nominal width of the
functions. An observation will always result in either a detection centred at
$x_{1}$ or at $x_{2}$, permanently collapsing the superposed wavefunction to
either $\psi(x-x_{1})$ or $\psi(x-x_{2})$. \emph{The gravity-induced collapse
mechanism invokes essentially the gravitational energy of a hypothetical
mutual interaction}, of matter distributions represented by the component
wavefunctions, as an instability measure that can induce such collapse.
\emph{The underlying logic is based on the literal or ontological
interpretation of a quantum superposition as spatially separated matter in
real space, as schematically indicated in the figure} \ref{super}. We stress
the caveat that such an interaction `between' the wavefunctions of a
superposition is purely hypothetical, requiring the (failed) identification of
the wavefunctions with the `fuzzy' matter distributions in real space, as was
attempted in the very early days of quantum mechanics. In any case, this
energy is identified by Penrose \cite{Pen1} to be the expression
\begin{equation}
\Delta E=4\pi G\int\left(  \Phi_{1}-\Phi_{2}\right)  \left(  \nabla^{2}%
\Phi_{1}-\nabla^{2}\Phi_{2}\right)  d^{3}x
\end{equation}
where $\Phi$ is the gravitational potential of each density distribution
corresponding to the individual wavefunction in the superposition. This can be
rewritten in terms of the mass density distributions, corresponding to the two
configurations that are notionally implied by the superposition, as
\begin{equation}
\Delta E=4\pi G\int\int\left[  \rho_{1}(x)-\rho_{2}(x)\right]  \left[
\frac{\rho_{1}(y)-\rho(y)}{\left\vert x-y\right\vert }\right]  d^{3}xd^{3}y
\end{equation}
The same expression is derived in the Di\'{o}si mechanism as well. This
gravitational interaction energy of the \emph{difference between the mass
distributions} of each of the two quasi-localized stationary distribution is
the sole decider of the average duration of stability of a superposition, in
the D-P mechanism. In the argument advocating the role of gravity in
state-vector reduction, Penrose viewed the macroscopic quantum superposition
of two differing mass distributions \ as unstable (analogous to an unstable
particle). Therefore, it was speculated that such a superposition would decay,
or collapse, into one of the many states of the superposition, in a
characteristic `quantum time scale' related to the gravitational self-energy
$\Delta E$. Then, the average time for spontaneous collapse is $\tau=h/\Delta
E$. The gravitational interaction energy for quasi-localized configurations
with separation $\left\vert x_{1}-x_{2}\right\vert \approx a$ is $\Delta
E\approx Gm^{2}/a$.

We can now discuss the central significant result in this paper, that we
reliably rule out any role of gravity or any other known interaction in the
collapse of the superpositions of quantum wavefunctions. We will prove the
simple and transparent result that any scheme in which the component
wavefunctions that are spatially separated have an energy of self interaction
is inconsistent and self-destructive.

Consider the commonly discussed D-P mechanism in which the superposition has
two equal wavefunction components that are separated by a distance of the
order of their sizes (figure \ref{super}). The gravitational energy even for
the case of a particle of size 1 micron is very small, $E\approx
Gm^{2}/a<10^{-19}$ eV, and it does not affect the observable quantum dynamics.
Then, the time scale of the collapse is $\tau=h/E\approx10^{5}$ s. But this
becomes less than 0.1 s for a particle 10 time larger, suggesting the
plausibility of the D-P mechanism.

However, the D-P mechanism assumes that only gravity can be effective in the
collapse of the wavefunction, excluding arbitrarily other fundamental
interactions. This restriction is needed because gravity is the only known
universal interaction, applicable to all quantum systems. Though it is not
clarified how to deal with the other interactions, when they are present, this
itself is not an irreparable deficiency of the D-P mechanism. If we take the
Einsteinian space-time justification advocated by Penrose, then one can try to
justify that gravity is the unique choice by virtue of its ability to alter
the space-time itself. However, if the particle is charged, as the case in
numerous experiments in the laboratory, \emph{the interaction energy of the
electromagnetism is as real as the interaction energy of gravitation}; it is
unavoidable and no mechanism can have one type of energy without the other.
While the electromagnetic energy is ineffective in the collapse of the
wavefunction, \emph{by assumption}, it is extremely important in the dynamics
of the particle! \emph{None of the proponents for the gravity aided collapse
ever considered this simple and natural possibility, and its destructive
consequence to the D-P mechanism of the spontaneous collapse of the
wavefunction}. To see its crucial nature, let us calculate the electromagnetic
energy in exactly the same situation as we considered earlier with only
gravity. The expression is exactly the same as in the equation, with $G$
replaced by $1/4\pi\varepsilon_{0}$, and the mass distributions replaced by
the charge distributions. At the localizable separation (figure \ref{super}),
$\left\vert x_{1}-x_{2}\right\vert \approx a$, the electromagnetic energy is
$\Delta E_{em}\approx e^{2}/4\pi\varepsilon_{0}a$. The ratio of the two
contributions is
\begin{equation}
\Delta E_{em}/\Delta E\approx\frac{e^{2}}{4\pi\varepsilon_{0}Gm^{2}}%
\end{equation}
which is \emph{independent of the details of the superposition}. Since the
quantities $\nabla^{2}\Phi_{i}$ are nonzero, the mutual forces are also
nonzero and large, unlike the gravitational forces. Also, the ratio of the
mutual forces from the two interactions are independent of the details and
evolution of the superposition. We can examine the routinely encountered
situation of trapped ions of mass $m\approx10^{-23}$ kg, in a quantum
superposition of localized stationary states separated by $0.1$ microns
\cite{ion-sup}. While the mutual gravitational force is an utterly negligible
$10^{-42}$ N, the mutual electromagnetic force is $10^{-14}$ N, an unbearably
enormous force for an ion of the tiny mass of $10^{-23}$ kg! The resulting
acceleration of $10^{9}$ m/s$^{2}$ would catapult the ion out of the
laboratory. It is obvious that proposals like the D-P mechanism that invoke
the gravitational self interaction to induce the collapse of the wavefunctions
in \ spatial superpositions are irretrievably ruled out by simple physical considerations.

Having decisively ruled out the D-P mechanism of the gravity-induced collapse
of the quantum superpositions, we can now make some general remarks about
other issues of principle in these models. One is the inconsistency of the
theoretical scheme, which remains somewhat hidden as long as one considers
only the superpositions involving only two-component wavefunctions, as the
proponents of such spontaneous collapse mechanisms often do. But, when there
are many components in the superposition, the very formalism implies a
nonlinear increase in the self energy, violating the fundamental conservation
constraints, which renders the proposal inconsistent. The problem arises
because gravity can be invoked only to collapse the wavefunction, while the
actual values of the probabilities of realization, through the Born's rule,
are not related at all to the gravitational physics. For definiteness, we
consider a particle with the mass $m$, and size $a$, represented by the
spherically distributed superposition with $N$ component wavefunctions, with
their locations centred at $\vec{r}_{i}$,
\begin{equation}
\Psi=\sum_{i=1}^{i=N}\frac{1}{\sqrt{N}}\psi_{i}\left(  \vec{r}-\vec{r}%
_{i}\right)  \label{multi}%
\end{equation}
In the case of the two-component superposition, the instability energy $\Delta
E$ is scales as $m^{2}$, with no indication of the number $N$ of the component
wavefunctions. But, with the more elaborate wavefunction in the expression
\ref{multi}, the situation is very different. Following Penrose \cite{Pen2},
we can calculate the gravitational energy that will trigger the collapse. For
$\left\vert r_{i}\right\vert \approx a$,
\begin{equation}
\Delta E>G\left(  N-1\right)  Nm^{2}/2\left\vert r_{i}\right\vert +\frac
{3N}{5}Gm^{2}/a
\end{equation}
This should be contrasted with the self energy when $\left\vert r_{i}%
\right\vert \ll a$, which is just $3Gm^{2}/5a$, which clearly proves the
severe violation of the conservation of energy, inherent in the D-P mechanism.
We have to rigorously keep in mind that \emph{there is only one material
particle}. \emph{The wavefunction, however complex and detailed, is but a
representation of the physical state of the particle, and not the particle
itself}. Unfortunately, this crucial fact has been overlooked by the
proponents of the gravitational collapse models, and as a consequence, these
self-destructing features were not noticed.

In summary, we have conclusively ruled out the widely discussed and actively
researched mechanisms of the gravity-induced collapse of the spatial
superposition of quantum wavefunctions, employing transparent first principle
calculations and comparison with the factual empirical situation. Our result
is entirely independent of the details of these collapse mechanisms, and it is
obtained by the simple device of just considering other common aspects of real
matter in dynamics, like the electric charge and its interaction energy. With
this result, it is ever more clear that the solution of the quantum
measurement problem and the problem of the resolution of spatial
superpositions are to be sought in a deeper level of dynamics and quantum
mechanics \cite{Unni-RQM}. What seems conclusive is that gravity is not the
cure for the primary malady of quantum mechanics.

\end{document}